\documentclass[11pt]{article}

\usepackage{natbib,graphicx,color,url,hyperref}
\usepackage{soul}
\usepackage{pdflscape}
\usepackage{graphicx}
\usepackage{colortbl}
\usepackage{tabularx,ragged2e}
\usepackage{booktabs}
\usepackage[utf8]{inputenc}
\usepackage{array}
\usepackage{hyphenat} 
\usepackage{amsfonts}            
\usepackage{amssymb}             
\usepackage{amsthm}              
\usepackage{amsmath}
\usepackage[letterpaper]{geometry}

\bibpunct{(}{)}{;}{a}{,}{;}

\title{Calling The Dead: Resilience In The WTC Communication Networks\thanks{This work was supported in part by the Office of Naval Research (ONR), Scalable Tools for Social Media Assessment, N00014-21-1-2229}}

\author{
  Selena M. Livas \thanks{The two authors made equal contributions and, therefore, are listed in alphabetical order}\thanks{Department of Sociology, University of California, Irvine} \and
  Scott Leo Renshaw\footnotemark[2]\thanks{Department of Software and Societal Systems, Carnegie Mellon University} \and
  Carter T. Butts\thanks{Departments of Sociology, Statistics, Computer Science, and EECS; University of California, Irvine; SSPA 2145, UCI, Irvine, CA 92697-5100; \texttt{buttsc@uci.edu}}
}     

\date{August 9, 2023}

\begin{document}

\maketitle

\begin{abstract}
Organizations in emergency settings must cope with various sources of disruption, most notably personnel loss.  Death, incapacitation, or isolation of individuals within an organizational communication network can impair information passing, coordination, and connectivity, and may drive maladaptive responses such as repeated attempts to contact lost personnel (``calling the dead'') that themselves consume scarce resources.  At the same time, organizations may respond to such disruption by reorganizing to restore function, a behavior that is fundamental to organizational resilience.  Here, we use empirically calibrated models of communication for 17 groups of responders to the World Trade Center Disaster to examine the impact of exogenous removal of personnel on communication activity and network resilience.  We find that removal of high-degree personnel and those in institutionally coordinative roles is particularly damaging to these organizations, with specialist responders being slower to adapt to losses.  However, all organizations show adaptations to disruption, in some cases becoming better connected and making more complete use of personnel relative to control after experiencing losses.\\[5pt]

\end{abstract}

\section{Introduction}

Individuals confronting an ongoing threat grapple with an array of challenges, needing to identify the threat, determine an appropriate response, and minimize the loss of life and property in the process. Disasters pose a significant test for human communities, often introducing additional obstacles that complicate their usual modes of organization. Such obstacles can range from the disruption of local radio communications \citep{kean:911commission:2011}, telephone communication failures \citep{mondal:ETDR:2021}, or the loss or incapacitation of key personnel, such as those with training or managerial authority, who are relied upon in crisis situations. Our study explores what happens when an individual in such a dynamic system goes ``radio silent,'' or becomes unable to respond. Using empirically-calibrated models of interpersonal communication during an unfolding disaster, we consider the impact of such node removal on aggregate communication networks, functionally relevant dynamics, and on the reorganization of the communication system in response to disruption.

The remainder of the paper is structured as follows.  We begin in Section~\ref{sec:background} by briefly reviewing relevant background on how communication systems respond to disruption, and on factors relating to maintenance or restoration of function in the face of personnel loss.  Our data and methods are described in Section~\ref{sec:methods}, with the results of our simulation study provided in Section~\ref{sec:results}.  Discussion of additional issues is included in Section~\ref{sec:discussion}, and Section~\ref{sec:conclusion} concludes the paper.

\section{Background} \label{sec:background}

Although the details vary depending on organizational type and task structure, organizational function typically requires the ability to coordinate the activities of its members \citep{galbraith:1977}.  This generally requires the ability to disseminate information to the members of the organization; the ability of individuals with interdependent tasks to communicate with each other and/or a mutual superior who can resolve conflicts \citep{thompson:bk:1967,krackhardt.carley:pr:1998}; and the ability of organizational members receiving or discovering information of broader importance to direct this to others who may need it \citep{cohen.et.al:asq:1972}.  Secondarily, communication keeps members oriented on organizational goals, directs their attention, facilitates situational awareness, and raises morale \citep{auf.der.heide:bk:1989}; thus, preventing individuals from becoming isolated can also be an important consideration \emph{per se}.  

In the face of disruptions such as personnel loss (``damage''), such capabilities may become threatened.  Broadly, it is useful to think of changes in organizational function under disruption in terms of distinct notions of \emph{robustness} and \emph{resilience}.\footnote{We note that these terms are not used consistently in the literature, and many studies of robustness (as we define it) employ the term ``resilience.''  Here, we refer to such studies in terms of our terminology.}  Here, we use the term ``robustness'' to reflect the capacity of an organizational system to \emph{maintain} function in the face of damage, while ``resilience'' reflects the capacity of such a system to \emph{restore} function lost due to damage.  Of these complementary concepts, robustness is the better understood, having been widely studied in the context of organizational and biological networks (see e.g. \citet{klau.weiskircher:ch:2005} for an introduction).  Studies of resilience, by contrast, have been hampered by the need to have access to dynamic models that can capture the reorganization of networks to damage.  As we describe below, we are here able to leverage a set of empirically calibrated models for communication dynamics \citep{renshaw:networksci:2023} to probe such reorganization, giving us the ability to speak to resilience \emph{per se}.   

Before turning to our specific approach, it is useful to motivate our study by briefly reviewing relevant prior work on robustness and resilience in organizational communication networks.  In the section that follows, we then describe the models we are using, and the simulation experiments we employ to examine the consequences of personnel loss.

\subsection*{Structural Change}

As noted, a considerable literature exists on network robustness. Typically, such studies begin with observed and/or simulated networks and remove nodes or edges, examining how functionally relevant properties are altered by these changes (known generically as ``attacks'').  \citet{bellingeri:fphy:2020} provides an overview of link and node removal studies on real networks, finding applications in diverse fields like biology, ecology, transport, infrastructure science, informatics, economics, and sociology. These studies use node and link removal to assess indicators of robustness -- measures of a complex system's ability to maintain function after loss of connectedness or group members. These studies often aim to understand the types of attacks that cause the most damage, typically related to vital links or nodes \citep{bellingeri:fphy:2020}. Just as networks vary in robustness to various nodal attacks, we might expect a similar variation in resilience (a question we probe below).

Many robustness studies focus on ``breaking'' networks, essentially limiting their ability to function and seeing which factors most effectively lead to fragmentation, as well as which factors increase network susceptibility. \citet{boldi:socinfomat:2011} found that social networks tend to experience less disconnection upon node attacks than web-graphs, but the most efficient removal strategy was related to what they termed ``label propagation.'' This technique involves iteratively labeling nodes based on their neighbors and identifying hub-structures that may or may not be high out-degree nodes. This finding resonates with other work by \citet{qi:appsci:2019} on optimal network disintegration patterns of multiplex networks. Previous research on the World Trade Center (WTC) communication network has aimed to understand its robustness under attack. Findings suggest that during the 9/11 events, the organizations involved in the 17 radio communication networks maintained connectivity through a relatively small number of coordinators. However, the reliance on these coordinators was context-dependent. As was discussed by \citet{fitzhugh.butts:sn:2021}, the structures found in the WTC radio communication networks are hub-dominated; while it makes them robust to random node removal, they are severely vulnerable to targeted attacks that implicate the hubs (``degree-targeted attacks'' \citep{fitzhugh.butts:sn:2021,klau.weiskircher:ch:2005}). Those who find themselves in these coordinative, hub-structural roles, are the most likely to have high degree -- through disconnecting the networks via these important actors, these types of attacks tend to be fairly effective at disrupting the network and limiting the ability for information transmission to occur in the absence of subsequent reconfiguration \citet{fitzhugh.butts:sn:2021}.

Comparative studies have also considered networks from other complex systems (e.g., metabolic networks and infrastructure systems).  These, too, exhibit varying degrees of robustness to different types of attacks. For instance, scale-free networks, like the World-Wide Web, have high tolerance for random attacks, but are extremely vulnerable to targeted high-degree node attacks \citep{albert:nature:2000,callaway:physrevlet:2000}. Social networks, in contrast, are much more robust than Random Erdos-Renyi graphs \citep{demeo:tcyb:2018}. A review of the literature by \citet{bellingeri:fphy:2020}
concluded that many social networks are ``robust yet fragile'' to attacks -- in that random node removal attacks do not dismantle the network, while highly targeted attacks (like malicious attacks) can often quickly and easily cripple a social network.  Motivated by this observation, here test various hub attacks against random attacks to understand how the ability to reorganize is impacted by the type of nodes removed. 

In related work, \citet{lv:chaosfrac:2022} found in designing a model for cascading failure that also considers activity overload when actors are randomly removed, that the ``dynamical behavior'' of the complex system is also a critical factor to consider for a network's robustness. They noted that networks with ``birth and death'' and regulatory dynamics are much more sensitive than biochemical or epidemic dynamics to their use of a sensitivity factor (which involved the ability of nodes in the network to resist the perturbation) \citep{lv:chaosfrac:2022}. \citet{fitzhugh.butts:sn:2021} also considered a form of dynamic robustness, in which they examined the impact of failures on forward connectivity under observed dynamics (as opposed to reorganization in response to damage); they found generally similar patterns of vulnerability to those seen in time-aggregated networks, but noted that this may or may not hold when the network can adapt to damage.  We will revisit that question in our study below.

While past robustness studies have yielded important insights, \citet{bellingeri:jsp:2018} notes that different node removal approaches can yield results that are sensitive to how both the target network and the attack are specified, potentially leading to misleading results if analyses are not prepared and interpreted carefully. For example, they found that studies focusing on individuals with ``binary-topological indicators'' like the largest connected component might identify ``false hub-nodes,'' which could merely be nodes with many weak links; put another way, disruptions to network structure may or may not have equal functional significance, and it is important to consider how function is maintained in the network (e.g., for diffusive systems, what is diffusing across the network and how diffusion occurs). For instance, in an epidemiological study of vaccination inoculation, identifying individuals in hub-roles might not be the most strategic choice for inoculation, despite what traditional analyses might naively suggest. This is because false hub-nodes with higher binary connectivity might have many links that are exceedingly weak and therefore limited contact time or low probability of infecting others \citep{bellingeri:jsp:2018,bellingeri:fphy:2020}. Similarly, time aggregation can exaggerate the potential for disruption due to hub removal, leading to misleading conclusions regarding the efficacy of such attacks for inhibiting diffusion \citep{butts:s:2009}.  It may also be important, particularly in dynamic settings, to consider nodes with particular roles or other characteristics (e.g., in the WTC case, nodes occupying institutionalized coordinative roles), whose removal may have different consequences than the removal of other network members.  Finally, it must be observed that robustness itself may not always be of primary substantive interest, in contexts where reconfiguration (and hence resilience) is possible.  As noted by \citet{butts.carley:jms:2007}, there can exist regimes in which organizations can repair damage indefinitely, so long as the intervals between attacks are long enough relative to repair times; this constant need for damage repair imposes a \emph{homeostasis cost,} which can affect optimal organizational structure.  In such settings, the cost of adaptation may be a more important theoretical target than the potential for catastrophic failure.  This motivates closer study of adaptation and resilience, a significant concern of this paper.

\subsubsection*{Robustness in the WTC Case}

The goal of this and prior papers investigating communication in the 9/11 disaster have been to understand how individuals maintain effective communication in conditions that make it extremely difficult - indeed, conditions in which the kinds of tasks with which individuals are faced are often unlike anything seen in normal circumstances.  Prior work on the WTC case has attempted to characterize the emergent hub-structures seen in the communication networks \citep{petrescu-prahova.butts:ijmed:2008,butts.et.al:jms:2007}, identify mechanisms driving responder communication \citep{butts:sm:2008}, understand the dynamics of hub formation formation (i.e., what kinds of social forces create the hub-structures seen in the observed networks) \citep{renshaw:networksci:2023}, and examine the robustness of the communication networks to node removal \citep{fitzhugh.butts:sn:2021}.  This study continues this line of research by using dynamic models to probe resilience of the WTC networks to personnel loss. 

As noted above, \citet{fitzhugh.butts:sn:2021} studied the WTC communication networks to understand their robustness to attack.  Their work highlights a contrast between communication strategies that rely heavily on centralization to maintain connectivity of actors in the organization (which would thus favor efficiency and robustness to random failure) but have the added effect of relying on a small number of key players (relative to a decentralized network). Their study employed several node removal strategies using both time-aggregated and time-sequenced networks. For both, they considered random failure in uniform random order, random failure of individuals in Institutionalized Coordinative Roles (ICRs), degree-targeted failure, and degree-targeted failure of ICRs for all of the 17 WTC radio networks.  In the time-aggregated case, the WTC organizations were found to be more impaired by random failure of ICRs than they were by removal of random actors; thus, individuals in institutionalized coordinative roles play a vital role maintaining connectivity.  By contrast,  degree-targeted failure was more effective at impairing the network than degree-targeted failure of ICRs, suggesting that ICR status, while important, is not as consequential as the degree of the actor, which is more indicative of hub structure. This follows prior work which found that while ICRs tend to be more likely to occupy hub roles, the majority of such roles emergent \citep{petrescu-prahova.butts:ijmed:2008}.

A second major part of their study, which partially motivates our approach, involved an analysis of the impact of node removal on observed event sequences, thus respecting the consequences of (observed) dynamics.  Taking the observed ordering of events as given, they considered the ``temporal unfolding'' \citep{bearman:ajs:2004} of the dynamic network and the potential ``pathways of transmission'' for information through time \citep{fitzhugh.butts:sn:2021}. They find broadly similar results to the aggregated case (e.g., the WTC networks were more robust to random failure than to random failure of ICRs, and more degraded by degree-targeted failure than degree-targeted ICR failure), though the difference between degree and degree/ICR attacks was no longer significant. Overall, their study demonstrates the ability to utilize robustness analysis to understand and detect network properties that may not be easily detected through more traditional measures \citep{freeman:socnet:1979, wasserman.faust:bk:1994}, but raises the question of what happens when the stricken organizations are able to dynamically reorganize their communication pattern in response to damage. For instance, one may hypothesize that hub removal will have reduced impact in this adaptive scenario, because new hubs may simply emerge to replace the old; by contrast, ICRs (which cannot be replaced) may take on greater significance in an adaptive setting.  To examine these questions, we must go beyond the observed patterns of communication in the WTC disaster, to instead consider the observed \emph{behavioral dynamics} governing the network and ask how these dynamics unfold when the network is damaged.  This takes us beyond the realm of robustness, and into the realm of resilience. 

\subsubsection*{Resilience Studies}

While robustness to attack has been studied in both human and non-human contexts, there have been far fewer studies on resilience in human networks. While robustness relates to functionality under attack, resilience is the idea of regaining functionality and reorganizing in response to disruption \citep{deBruijne2010}. Social science research in this area has to date tended to focus less on understanding how systems dynamically respond to disruption, instead involving after-action case studies of how organizations and their members responded to major organizational failures. The organizational literature has had some focus on issues relating to resilience, particularly in the area of organizational learning; this includes e.g. work on the kinds of organizational structures that have been found to be more or less efficient in their ability to learn/respond to changes in their environment. Several simulation studies by Carley \citeyear{carley1991designing,carley:os:1992} provide context to organizational learning within settings of communication breakdown and or crisis. This research also provides insights in the context of turnover, i.e. organizational actor loss. Research by \citet{virany1992executive} found that executive turnover (leadership) may be necessary in a turbulent environment for organizations to adapt to changing circumstances -- the logic being that executives do not have training in the new crisis/turbulent environment and that prior organizational scripts and routines might be counter-productive and even detrimental to organizational success. Having some turnover might allow for organizations to be more flexible overall. While node removal is generally viewed as a disruptive event, this work suggests the intriguing possibility that it may not always lead to impairment, and indeed that it could in some cases actually improve organizational performance.  The idea that organizations can dynamically ``build back better'' in emergency settings is counterintuitive; as we show, however, some such \emph{hormetic} or \emph{eustretic} behavior is seen in the case of the WTC networks.

\subsection*{Emergencies and Efficient Communication}

Effective communication for coordination between organizational members in nominal or disrupted context is critical, with early research finding that effective patterns of communication are ``a first principle of effective performance" \citep[p.594]{bavelas:jacsa:1950}. More generally, prior work has found that successful organizations and effective communication between organizational members are critical to many performance measures \citep{chandler:1977, galbraith:1977, malone:ms:1987, minsky:1986,wageman:ASQ:1995, brusoni:adminsciquart:2001}, and that effective coordination in tasks enhances group performance \citep{tushman:ASQ:1979, liang:PSPB:1995, dedreu:PNAS:2016}. To ensure that information can flow optimally, organizations need to utilize reliable channels to collect and disseminate their information. When these important channels fail the consequences can result in many counterproductive effects, including task delay or even failure to execute \citep{chandler:1977, galbraith:1977, kim:CMOT:2002, radner:JEL:1992, sah:AER:1986, tjosvold:HR:1984, krackhardt:TOB:1996}. 

Thus, one of the main areas in which communication has been studied is in the context of emergency communications -- events that can often lead to communication interruptions or failures in a context where information can be extremely time-sensitive. In their review of emergency communication and the use of Information and Communications Technologies (ICT) in disaster management, \citet{mondal:ETDR:2021} summarizes the impact of several large-scale natural disasters over the past decade. They report that disasters such as floods, hurricanes, and earthquakes affected television and cell phone communications for a duration ranging from 3 or 4 days to more than three months, as seen in 2020 during Hurricane Maria in Puerto Rico. In scenarios where modern connectivity options failed, they observed that amateur radio, in addition to portable satellite equipment, was employed \citep{mondal:ETDR:2021}. Radio communications, including amateur radio communications, has been a critical medium, particularly during the contexts of disasters -- there are studies that find that as a decentralized communication medium, it can weather the impact of disasters more effectively than highly centralized communication infrastructure like cell phone towers \citep{gill:IJES:2020,nollet:TAS:2013,colie:DPAM:1997}.

While radio communication is a decentralized medium, centralization of communication can still occur via a process in which coordinative and organizational responsibilities become concentrated around a relatively small number of individuals. Such centralized communication structures and groups have been found to exhibit advantages in executing decomposable tasks, resulting in better performance \citep{hage:amersocirev:1979, tushman:ASQ:1979, brusoni:adminsciquart:2001}. In the classic Bavelas task-performance and structure study, it was found that task groups with high, localized centralization were capable of learning more quickly, had fewer errors in their performance, and tended to be more stable than groups with low centralization \citep{bavelas:jacsa:1950}. However, it has also been noted that centralization, a kind of structural adaption, can have drawbacks including the inability to perform effectively in the context of non-decomposable tasks, like multilateral negotiation \citep{carley:os:1992}.

While structure has been a focus of the early literature, more recent studies have found that what is ``optimal'' or more ``efficient'' for communication is not merely determined by structure, and that for certain tasks, environments, or organizational contexts, there may be advantages or disadvantages to distinct structural forms. Despite early research pointing to ``ideal-types,'' there is likely a menagerie of structural forms that may be ideal depending on the context \citep{scott:1981,levitt:MS:1999,donaldson:2001,shenhar:MS:2001}. 

Several studies have found that organizations leverage role differentiation and task routinization through specialization that allows for coordination among interdependent roles and sets of individuals with particular skills and abilities \citep{hage:amersocirev:1979, kogut:orgsci:1996, vandeven:amsocrev:1976, brusoni:adminsciquart:2001, crawford:acadmanrev:2013}. Prior research has found that specialized roles and their associated routinization of tasks can help with efficiency of task performance \citep{cohen:ap:1994,kalleberg:ambehavsci:1994}. Consolidating tasks to more particular, specialized, roles, has been found to enhance emergency response organizations' information flow \citep{comfort:publicadminrev:2007} -- also noting that while these roles are well designed and operational in nominal/ordinary circumstances, it can often lead to rigidity, poor performance, and an inability to adapt in the face of tasks and contexts that those groups have had little experience with or preparation for \citep{comfort:publicadminrev:2007}. \citet{comfort:publicadminrev:2007} discusses that organizations might need to design role structures, i.e., specialization of roles, with the consideration that they may need to be adapted in contexts with limited resources, time, and more novel tasks.

To this point, \citeauthor{carley:ABS:1997}'s (1991) overview of organizational learning in the context of actual hazard and disaster responses found that training organizational members on standard operating procedure may not necessarily degrade performance, but that leaning too much on standard operating procedure can lead to organizational rigidity where personnel follow things dogmatically -- they found that newer personnel at lower organizational levels (less-specialization) tended to follow SOP unless they encountered new situations where the procedures did not cleanly apply -- they then fell back on personal experiences which can fruitfully lead to creative solutions. However, mid-level, more specialized personnel tended to follow SOP to the point of rigidity, not seeing past their organizational script - which in catastrophic disaster events, due to their rare nature are difficult to train personnel for, and learning is hard to transfer. They found that teams where personnel were empowered to act on the basis of their experience tended to outperform teams that strictly followed SOP. While specialized networks are trained to respond to crises, this might actually decrease their ability to respond to changing circumstances and we might suspect resilience to vary by the specialization of our networks.
 
The WTC networks also vary by their level of specialization -- some of the organizational units were trained to respond to crises and others were not. In the static case, \citet{fitzhugh.butts:sn:2021} looked at the difference between specialist and non-specialist networks. When they teased apart the specialist/non-specialist difference, they found that random failure of ICRs was significantly more devastating to network structure connectivity than random failure, but only for specialists; non-specialists seem to be robust to random ICR removal. They explain that this may be that these institutionalized coordinator roles plays a particularly important consideration in specialist networks, where nonspecialists tend to be less reliant on these individuals.

Organizations in environments with high levels of uncertainty around tasks, or at least novelty of tasks, may be better suited for decentralized teams, which can operate more efficiently than a hierarchy could, saving cost and time \citep{kim:CMOT:2002,vandeven:amsocrev:1976}. Early theoretical work has provided some inclination that teams may in fact learn faster than hierarchies \citep{carley:os:1992}. With individuals embedded in a hierarchy, more centralized, key individuals, may be subject to more information overload in these novel, uncertain, contexts; making them less effective in these contexts \citep{carley:os:1992}. It has also been found that novel crisis situations give way to more decentralized communication patterns among organization members \citep{uddin:computmathorgtheory:2011}, and that adaptation to crisis can ultimately be more efficient in the absence of centralized forms of communication \citep{pitt:computmathorgtheory:2011,rodan:computmathorgtheory:2008}.

\subsection*{Functionality and Resilience}

In order to understand better what mechanisms might help maintain functionality under pressure, we draw on studies related to social insects, which have used, in more recent years, a much more systematic approach to the issue of systemic resilience. Like human societies, social insects have complex social interactions that greatly influence group-level fitness, information sharing, and cooperation \citep{easter:royalsoc:2022}. The field of entomology has a rich history of studying these social dynamics among various social insect species \citep{wilson:bk:1971,holldobler:bk:2009}. One area of social insect research involves experimental and simulation studies that manipulate some aspect of the colonies' environment or the colonies themselves to understand the resulting dynamic changes. For example, some studies have increased the temperature in beehives \citep{johnson:bes:2002,cook:anibeh:2013} or augmented food availability \citep{pasteels:bk:1987,beckers:insectsoc:1990,traniello:bk:1995} to induce changes in the colonies' behavioral dynamics. Relevant to our current study are the removal studies, where researchers study worker loss either through experimental approaches (e.g., physically removing ants) or in-silico simulations to understand how the colony dynamics adjust to the loss of workers. Several studies have shown that the workforce is usually replaced by other workers in the colony \citep{johnson:bes:2002, mirenda:anibeh:1981, huang:bes:1996,gordon:anibeh:1986, wilson:bes:1983,wilson:bes:1984, mcdonald:jcp:1985, charbonneau:ICB:2017}, thereby maintaining functionality despite the removal of actors. 

While broadly the idea of ``social loafing,'' described as the tendency for individuals to lower their productivity when participating in a larger group \citep{ringelmann1913,ingham1974ringelmann,simms2014social}, has a negative connotation (often as a social disease \citep[p. 831]{latane1979many} some have speculated that, in social contexts, it could be an adaptive strategy relating to the conservation of resources \citep{williams1991social,bluhm2009adaptive}. While there has not been much further engagement with the positive re-framing of ``social loafing,'' similar contexts have found that reserves of actors and organizational slack (i.e, individuals who are not actively engaging but may be mobilized) may play an evolutionary advantageous role in certain systems. Surprisingly, in the case of social insects, it has been found that up to 50\% of workers in several species of social insects are inactive at any given time \citep{charbonneau:ICB:2017}. This has been observed in honey bees \citep{lindauer:jcp:1952, moore:jip:2001, moore:BE:1998}, bumble bees \citep{jandt:IS:2012}, wasps \citep{gadagkar:ZFPT:1984}, termites \citep{maistrello:IS:1999}, and ants \citep{charbonneau:BE:2015, herbers:psyche:1983, cole:behavecolsociobiol:1986, dornhaus:PLoSBio:2008}. It had been commonly hypothesized that this worker ``reserve'' can be mobilized quickly if workload increases \citep{charbonneau:PLoSONE:2017}.

\citet{charbonneau:PLoSONE:2017} tested this hypothesis by systematically removing highly active workers, inactive ants, and randomly selected workers from a colony of \textit{Temnothorax rugatulus} ants to see how the overall activity of the colony was affected. In support of the commonly held hypothesis, they found that the colony was able to maintain pre-removal activity levels when highly active workers were removed, supporting the idea that inactive workers serve as a pool of ``reserve'' labor. Interestingly, when inactive workers were removed, the level of inactivity decreased and remained lower after the removal period, suggesting a system-level ability to maintain active worker levels but not specific inactive worker levels \citep{charbonneau:PLoSONE:2017}.

Other social insect research has demonstrated that during heat stress, an environmental emergency, honey bees have been observed to switch tasks more frequently and prioritize specific tasks that aid in thermoregulation across all types of reserves, not just specific worker groups \citep{johnson:bes:2002}. Theoretical evidence suggests that this ability to reallocate reserves and workers based on task-switching could support colony survival during major catastrophes or large-scale disturbances to worker populations \citep{hasegawa:SciRep:2016}. This body of research indicates that social systems have mechanisms in place to respond to and regulate worker loss, aiming to maintain average work activity and preserve the functionality of the social group in both standard and emergency contexts.

Finally, within organizational contexts, there has been a line of work that has looked at how organizations responded to environmental shifts and to better understand which organizations tend to be most resilient/robust to these exogenous shocks/shifts through their use of ``slack resources" \citep{cheng1997organizational}. Bourgeois \citeyear{bourgeois1981measurement} discussed organizational slack as ``a cushion of excess resources.'' Researchers argued that slack provides the ability for the flexible use of resources, to be mobilized in uncertain contexts / new environments to adapt and enhance an organizations ability to respond to these new contexts \citep{carter1971behavioral,cyert1963behavioral,mohr1969determinants}. Often in this literature organizational slack is operationalized as excess financial resources that can be leveraged in a pinch, however slack could also include humans as a resource to be mobilized as well. 

\citet{charbonneau:PLoSONE:2017} offers an overview of how other complex social contexts could benefit from a systematic approach such as that used to study social insects. They discuss the potential that human organizations might possess fewer reserves, as humans typically operate in more predictable environments than social insects, or that human organizations might not adequately account for variability and should consider more flexible workers to optimally navigate diverse environments \citep[p.15]{charbonneau:PLoSONE:2017}. Through our current study, we hope to contribute to this larger research on social systems by providing insights into the potential flexibility of humans in situations where actor incapacitation or loss might positively or negative impact the functionality of the collective system. Our study leverages the simulation capabilities in the \texttt{relevent} \textsf{R} package \citep{butts:relevent:2013}, to mobilize a battery of simulation studies to test the impact of various forms of node removal attacks on measures of resilience, building off prior work to find in which ways these networks are resilient. 

\section{Data and Methods} \label{sec:methods}

The data we use for this study comes from coded transcripts of radio communications among specialist and non-specialist responders during the World Trade Center disaster on the morning of September 11th, 2001. This data was extracted from transcripts released by the Port Authority of New York and New Jersey; they were originally coded by \citep{butts.et.al:jms:2007} and were recently made publicly available \citep{renshaw:networksci:2023,butts.et.al:wtc:2021}. Specifically, these data are from seventeen organizational units associated with the Port Authority of New York and New Jersey, each of which was communicating internally using handheld radios.  The data consists of the sequences of radio calls among named communicants within each group, beginning when the first plane crashed into the WTC North Tower at 8:46 am, and extending for 3 hours and 33 minutes or until communication was terminated by structural collapse (in the case of some groups who were inside the WTC complex). Further background on the data set can be found in \citet{butts.et.al:jms:2007,petrescu:ijmed:2008}.

In prior work, \citet{renshaw:networksci:2023} used relational event models (REMs) to examine the communication dynamics among WTC responders, with a particular eye to identifying mechanisms responsible for hub formation. They generated best-fitting relational event models \citep{butts:sm:2008} for each of the 17 radio communication networks, using AICc-based model selection criteria and validation via simulation-based model adequacy checks. We leverage these empirically calibrated models for the purposes of this simulation study.  Our general approach is as follows.  In each simulation replicate, we perform the following for each network:
\begin{enumerate}
\item We fix the first half of the observed event history (starting our simulation at the half-way mark).
\item We identify a set of nodes (chosen by an \emph{attack mechanism,} as described below) to be incapacitated; these are ``marked,'' and rendered incapable of sending messages.
\item Using the empirically calibrated model for the specified network (subject to the additional constrained that removed actors cannot send), we simulate 600 additional events conditional on the history in (1).
\end{enumerate}
The resulting simulated sequences are then analyzed to examine the impact of the attack on communication network structure and dynamics; simulated sequences in which no individuals were incapacitated were used as controls.  As described below, we analyze the ability of the WTC networks to respond to different types of personnel loss, allowing us to probe the resilience of the communication system; importantly, while the \emph{tendencies} of agents within each network are held to their empirically calibrated values, their actual behavior is free to adapt as the network evolves.  This therefore complements the traditional network robustness studies reviewed in Section~\ref{sec:background}, in which observed communication patterns are held fixed (net of vertex removal).

\subsection*{Simulation Design}

Our procedure is implemented as follows.  We first take the empirically calibrated parameters for each model reported by \citet{renshaw:networksci:2023} and generate a corresponding model skeleton using the \texttt{relevent} package \citep{butts:relevent:2013} in the \textsf{R} statistical programming language \citep{rcore:sw:2020}.
Each model skeleton includes a covariate for whether a given node occupied an ICR, as well as a binary covariate indicating the nodes to be incapacitated; a sender covariate effect was added to the calibrated model for the incapacitation covariate, with a sufficiently large (i.e., effectively numerically infinite) negative coefficient to ensure that the hazard for sending from incapacitated nodes was numerically zero.  (The values of the incapacitation covariate were set using the appropriate attack mechanism, as described below, with control cases having no incapacitated nodes.)  Using the \texttt{relevent} simulate function, we then simulate additional events from these models, fixing the first half of the event history to the empirical data, generating 600 new events into the future after node removal. For instance, our largest network, Lincoln Tunnel, has a total of 1146 events in the empirical network; we thus use the first 573 events as the starting point for our simulated sequences, simulating 600 events past this point for a total of 1719 events.  Our study hence employs an \emph{in silico} interrupted time series design, where we compare \emph{treatment} histories in which nodes were incapacitated at a specific point to \emph{control} histories in which the intervention was not performed, with both prior history and behavioral mechanisms held constant. 

With this as our base framework to simulate new event sequences, we then consider network evolution under several attack mechanisms; a descriptive table of node attacks with a breakdown of the number of simulations in each condition can be found in Table \ref{tab:simulation_counts}. We start out by creating a comparative baseline (i.e., control) condition for each of the 17 communication networks, where no attacks were performed. In these baselines, the first $n$ events are still fixed, with $n$ being the halfway point of the empirical networks. With no nodes removed from the network, we then simulate 600 events into the future. We conducted 100 independent simulations for the control condition for each of the 17 networks. We then simulated network evolution under four different attack mechanisms: Degree Attack, ICR Attack, Combined Attack (i.e., Degree and ICR), and a Random Attack. The degree attacks were constructed by sorting the actors by their total call volume (valued Freeman degree) in the empirically observed networks and then removing the highest k\% of nodes. For ICR Attack, we randomly ordered all nodes occupying ICRs, followed by a random ordering of all non-ICR nodes, and took the top k\% of nodes in this sequence (i.e., selecting ICR nodes first). For the Combined network attack, we first randomly ordered our ICR actors, and then ordered the non-ICRs from highest to lowest empirical degree; we then removed the first k\% of actors (i.e., first removing ICRs at random, and then removing remaining nodes in descending degree order). Finally, Random Attack was conducted by randomly selecting k\% of actors to be removed regardless of degree or ICR status. We conducted each of these attacks with $k$ set to 5\%, 10\%, 15\%, 25\%, and 50\%, for a total of 20 different attack mechanisms -- generating a new node removal vector for each simulation within each attack. For each of these 20 attacks, we generated 100 different simulations per network, for a total of 35700 simulated event sequences (including our 1700 control simulations). 

We note that, unlike traditional robustness tests that simply remove nodes from a fixed communication network, our protocol allows incapacitated nodes to remain \emph{targets} of communication.  The simulated actors in the network do not initially ``know'' that their alters have been removed, and indeed may spend time and resources attempting to contact them (the titular act of ``calling the dead''). This is similar to how real incapacitation or death during a disaster often occurs, with radio silence from the person on the other end, and is reflective of the the environment of the WTC response in which agents typically had to infer who was present and active from observed communication activity \citep{butts.et.al:jms:2007}.  By observing the rate at which organizations are able to detect losses and reconfigure communication in response to them, we are able to probe their resilience to personnel loss.  Further, by examining the variation in resilience across attack mechanisms, we are able to probe the relative dependence of each organization on high degree versus institutionally defined coordinators.

\begin{table}[h]
\centering
\resizebox{0.8\textwidth}{!}{%
\begin{tabular}{l|*{6}{c}}
\hline
 & Baseline & Degree & ICR & Combined (Degree \& ICR) & Random & Total \\
\hline
Baseline & 1,700 & - & - & - & - & 1,700 \\
5\% & - & 1,700 & 1,700 & 1,700 & 1,700 & 6,800 \\
10\% & - & 1,700 & 1,700 & 1,700 & 1,700 & 6,800 \\
15\% & - & 1,700 & 1,700 & 1,700 & 1,700 & 6,800 \\
25\% & - & 1,700 & 1,700 & 1,700 & 1,700 & 6,800 \\
50\% & - & 1,700 & 1,700 & 1,700 & 1,700 & 6,800 \\
\hline
Total & 1,700 & 8,500 & 8,500 & 8,500 & 8,500 & 36,700 \\
\hline
\end{tabular}%
}
\caption{Number of observations for different node removal strategies and removal percentages, with totals. \label{tab:simulation_counts}}
\end{table}

\subsection*{Outcome Measures}

In evaluating our simulated trajectories, we focus on three categories of measures: structural changes, efficiency and functionality, and the role of reserves. Within each category we test for differences in the type of attack (combined, degree, ICR, and random), the percentage of nodes removed (5\%, 10\%, 15\%, 25\%, and 50\%), and the differences between specialist and non-specialist networks. In order to evaluate differences in outcomes we use a series of $t$-tests, comparing networks under node-removal conditions and under baseline (i.e., treatment vs. control).

We use the following five measures to evaluate treatment effects on aggregate network structure.  The first is the Theil index of communication volume, which is used as a measure of hub formation; an inequality measure, the Theil index is higher in networks with greater hub structure (a major feature of the WTC networks).  This measure replicates that used in \citet{renshaw:networksci:2023}. Next we measure the Krackhardt connectedness \citep{krackhardt:ch:1994} of the aggregated network, in parallel to robustness studies that frequently focus on fragmentation as a result of node removal. Our third measure is the degree centralization of the aggregated network, employed to test how evenly distributed communication volume is across each network; centralization has also been used frequently in studies of network efficiency. We also measured aggregated graph density to understand how many potential connections were realized. Lastly, we measured the proportion of the graph consisting of isolates (nodes that do not send or receive communications) in order to get at how much of the network is involved in communicating and how much slack is present under each condition. Taken together, we can understand whether or not these networks are fragmenting after node removal, or if they display a different pattern of resilience.

To measure efficiency and functionality, we measure the number of calls directed to incapacitated nodes and the average forward reachability within each simulated network. More efficient networks will waste fewer calls on those who cannot respond and learn more quickly not to direct calls to them.  Likewise, communication networks only function if they are able to disseminate information, motivating the fraction of forward-reachable pairs as a measure of functional capacity. 

Finally, the literatures on network resilience, slack resources, and organizational learning all point to the role of reserves in maintaining network functionality and responding to crisis. We thus construct a measure of reserve use (i.e., mobilization of previously inactive agents) and compare this across network conditions to understand their role in the reorganization and functionality of our radio communication networks. 

\section{Results} \label{sec:results}

\subsection{Changes In Network Structure}

\subsubsection*{Attacks Induce Coalescence}

\begin{figure}
\begin{center}
\includegraphics[width=0.9\textwidth]{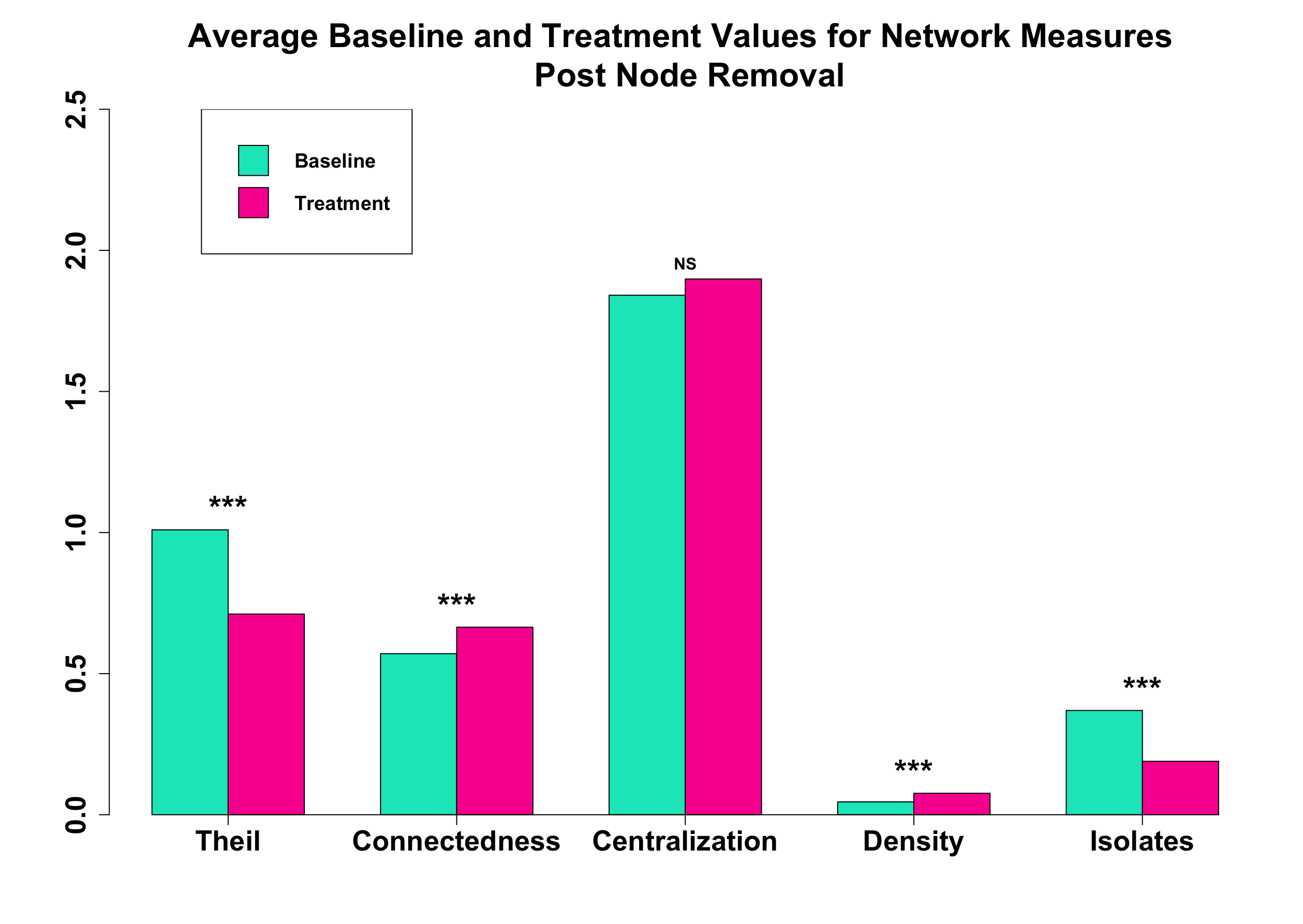}
\caption{Mean values for time-aggregated network properties over all networks for treatment and control simulations ($*** = p<0.001$).  Overall, attacks lead to enhanced cohesion versus baseline simulations. \label{fig:smryBars}}
\end{center}
\end{figure}

In order to capture the pattern of network resilience, we focus on the five general measures from Section~\ref{sec:methods}: the Theil index, network connectedness \citep{krackhardt:ch:1994}, degree centralization \citep{freeman:socnet:1979}, density, and proportion of isolates. These measures allow us to understand how the networks come together, or fragment after node removal, how much of the network is involved in communication, and how concentrated communication activity is. Figure~\ref{fig:smryBars} shows the average measure across all 17 networks after node removal compared to the average measure in the control condition, in which no nodes were removed. All measures are taken on the network over the 600 simulated events (i.e., not including the pre-history) in order to make all measures comparable. Rather than fragmenting, these communication networks appear to show a general trend of coalescence: density and connectedness are higher, while the Theil index, and the proportion of isolates are lower compared to the baseline simulations, while centralization remains unchanged. This implies that these networks have more actors involved, more connections between the actors, and are less hub-dominated following node removal. 

\begin{figure}
\begin{center}
\includegraphics[width=0.9\textwidth]{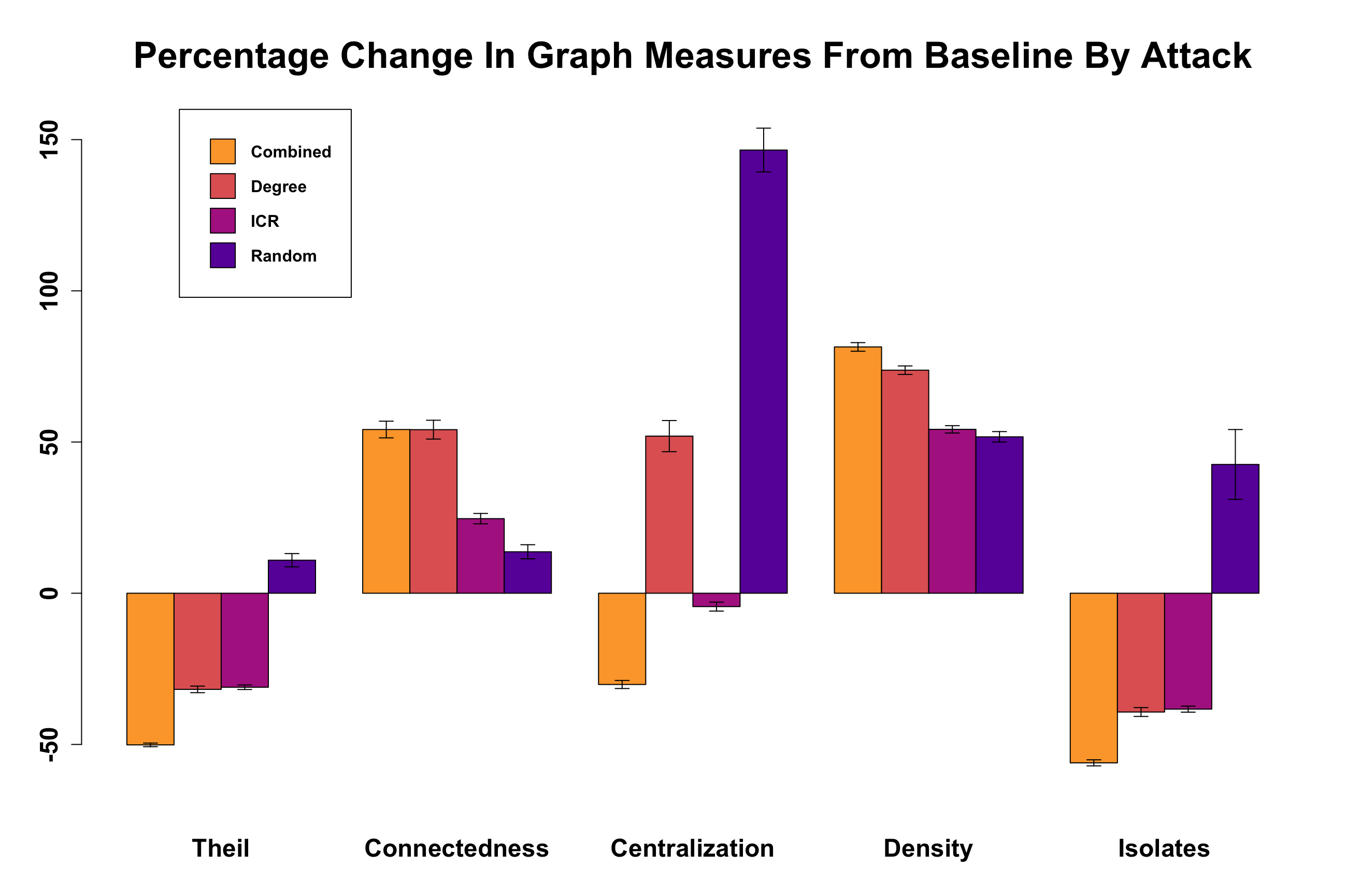}
\caption{Percentage difference in time-aggregated network properties versus control over all networks, by attack mechanism.  Cohesion enhancement is broadly consistent over targeted attacks, but not for random node removal. \label{fig:change.atkCI}}
\end{center}
\end{figure}

We also tested whether or not this result varied by attack type and percentage of the network removed. Figure~\ref{fig:change.atkCI} shows mean treatment/control differences by attack mechanism. All measures of interest have statistically significant differences in means between all categories except for Theil index, where ICR and degree attacks are not statistically different, connectedness, where degree and combined attacks are not different, and proportion of isolates, where ICR and degree and not different on average. Overall, we can see that attack types do tend to vary from one another in their impacts on changes from the baseline network structure, but that all targeted attacks tend to enhance cohesion. We can also see that random attacks behave quite differently than targeted attacks, increasing Theil index, centralization, and the proportion of isolates, while increasing density and connectedness less than the targeted attack types. The only other attack type that shows a contradictory effect is degree attacks on degree centralization, but we suspect that this may be due to the nature of degree centralization -- if there are many high degree nodes, targeting a fraction of them could actually result in a network of less active individuals and a smaller subset of very active nodes, thereby increasing our centralization measure.

\begin{figure}
\begin{center}
\includegraphics[width=0.9\textwidth]{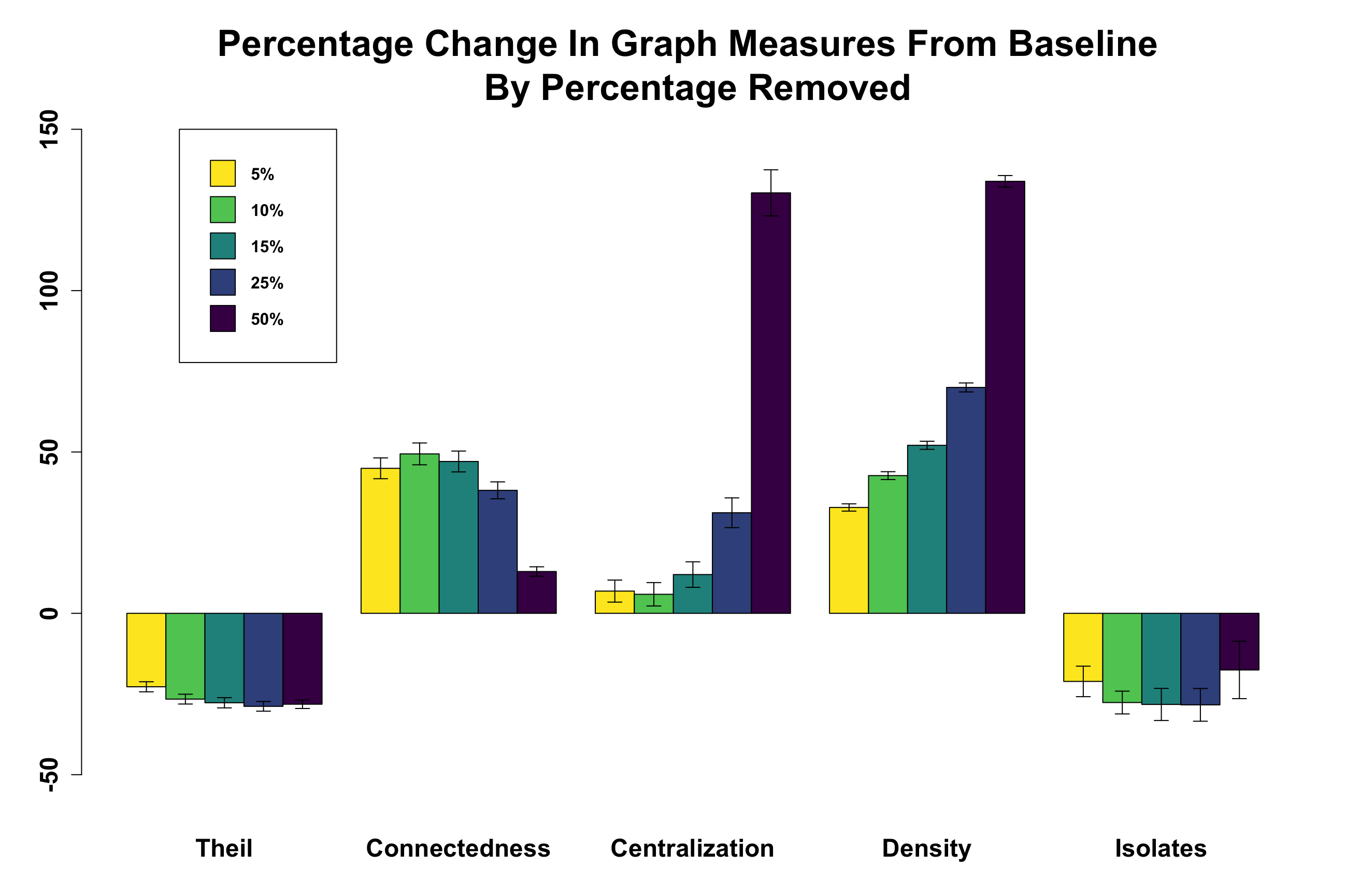}
\caption{Percentage difference in time-aggregated network properties versus control over all networks, by fraction of nodes removed.  Removal fraction has non-monotone effects for all outcomes other than network density, with the strongest effects observed at moderate levels of node removal. \label{fig:change.percCI}}
\end{center}
\end{figure}

We conducted the same test for percentage of nodes removed and found that there are more similarities in changes from baseline by percent than by attack. The only category that shows starkly different behavior is 50\% node removal, likely due to the extreme nature of removing half of the network. For the other 4 categories of removal, there is a general pattern of increasing coalescence up to approximately 10\% to 15\%, with 25\% removal producing similar changes in Theil index and isolate reduction, but producing slightly more centralized, less connected, and denser networks compared to 15\%. Overall, we find that both the attack type and amount removed will impact the resulting re-structuring of the network, but in general targeted attacks that remove less than 50\% of the network produce a pattern of \emph{coalescence,} rather than the fragmentation expected from robustness studies.

\subsubsection*{Specialist and Non-Specialist Networks Respond Differently}

While our general findings appear in line with the literature on node removal in both insect studies and social communication networks, we also consider whether there is substantial variation in the effects of removal across different types of networks. The main grouping of interest in the WTC case is networks of specialist versus non-specialist responders, corresponding to whether the organizational unit is one that would be trained and organized to respond to emergency situations as part of their normal repertoire. The networks are divided fairly evenly between these categories, with 9 specialist networks and 8 non-specialist networks. We thus disaggregate the findings from above and test for differences between the two groups. 

When we group the changes by node removal percentage, we can see higher levels of coalescence across all groups, with the only statistical difference being in isolate reduction at 50\% removal. Looking at attack type, we can see that in general, non-specialist groups tend to show greater magnitudes of coalescence with higher decreases in hub structure and isolate percentage, and larger increases in density. However, when we look at ICR attacks, we see that specialists networks show the largest decrease in both hub structure and isolate percentage. This may indicate that the role of ICRs differs between the two groups. It may be that non-specialist networks have more non-ICR emergent coordinators compared to specialist networks and that the resilience behavior we observe is driven by the removal of coordinator nodes, which may also explain the distinct effect of random attacks.

\begin{landscape}

\vspace*{\fill}
\begin{table}[ht]
\centering
\scalebox{0.78}{
\begin{tabular}{rrrrrrrrrrrrrrrr}
  \hline
  &\multicolumn{3}{c}{Theil} & \multicolumn{3}{c}{Connectedness}& \multicolumn{3}{c}{Centralization}& \multicolumn{3}{c}{Density}& \multicolumn{3}{c}{Isolates} \vspace{5pt}\\
 & Specialist & Non-Spec.& Diff & Specialist & Non-Spec. & Diff & Specialist & Non-Spec. & Diff & Specialist & Non-Spec. & Diff & Specialist & Non-Spec.& Diff \\ 
  \hline
Combined & -50.30 & -49.98 && 24.83 & 87.11 &***& -31.36 & -28.84 && 70.50 & 93.76 &***& -58.56 & -53.30&*** \\ 
  Degree & -23.89 & -40.68 &***& 20.73 & 91.63 &***& 94.50 & 4.09 &***& 58.41 & 91.03 &***& -33.01 & -46.33 &*** \\ 
  ICR & -43.44 & -30.82 &***& 20.06 & 40.08 &***& -19.16 & 9.34 &***& 59.01 & 71.79 &***& -52.41 & -38.84 &*** \\ 
  Random & 29.20 & -6.92 &***& -1.58 & 36.33 &***& 165.38 & 96.53 &***& 31.65 & 59.83 &***& 87.25 & -6.61 &*** \\ 
   \hline
\end{tabular}
}
\caption{Mean change from baseline for five structural measures by attack type and specialization. The difference between specialist networks and non-specialist networks was tested  ($*** = p < 0.001$, $** = p < 0.01$, $* = p < 0.05$). Specialist networks tend to show less coalescence across all attack types expect for ICR attacks.\label{tab:specNonDiffs_AtkStrucMsres}}
\end{table}

\begin{table}[ht]
\centering
\scalebox{0.78}{
\begin{tabular}{rrrlrrlrrlrrlrrl}
  \hline
  &\multicolumn{3}{c}{Theil} & \multicolumn{3}{c}{Connectedness}& \multicolumn{3}{c}{Centralization}& \multicolumn{3}{c}{Density}& \multicolumn{3}{c}{Isolates} \vspace{5pt}\\
 & Specialist & Non-Spec. & Diff & Specialist & Non-Spec. & Diff & Specialist & Non-Spec. & Diff & Specialist & Non-Spec. & Diff & Specialist & Non-Spec. & Diff \\ 
  \hline
5\% & -15.97 & -30.39 & *** & 15.97 & 77.53 & *** & 17.37 & -4.91 & *** & 21.13 & 45.95 & *** & -4.7 & -39.54 & *** \\ 
  10\% & -19.63 & -34.4 & *** & 17.79 & 84.98 & *** & 22.58 & -12.87 & *** & 31.51 & 55.24 & *** & -12.6 & -44.56 & *** \\ 
  15\% & -22.74 & -33.3 & *** & 18.56 & 79.11 & *** & 27.09 & -4.94 & *** & 41.85 & 63.56 & *** & -17.8 & -39.98 & *** \\ 
  25\% & -25.47 & -32.6 & *** & 17.76 & 61 & *** & 46.9 & 13.46 & *** & 59.69 & 81.61 & *** & -20.36 & -37.36 & ** \\ 
  50\% & -26.72 & -29.8 & * & 9.98 & 16.31 & *** & 147.76 & 110.65 & *** & 120.29 & 149.15 & *** & -15.46 & -19.9 &  \\  
   \hline
\end{tabular}
}
\caption{Mean change from baseline for five structural measures by percentage removed and specialization. The difference between specialist networks and non-specialist networks was tested  ($*** = p < 0.001$, $** = p < 0.01$, $* = p < 0.05$). Specialist networks tend to show less coalescence regardless of percentage removed.\label{tab:specNonDiffs_PercStrucMsres}}
\end{table}
\vspace*{\fill}
\end{landscape}

\subsection{Network Functionality}

\begin{figure}
\begin{center}
\includegraphics[width=0.9\textwidth]{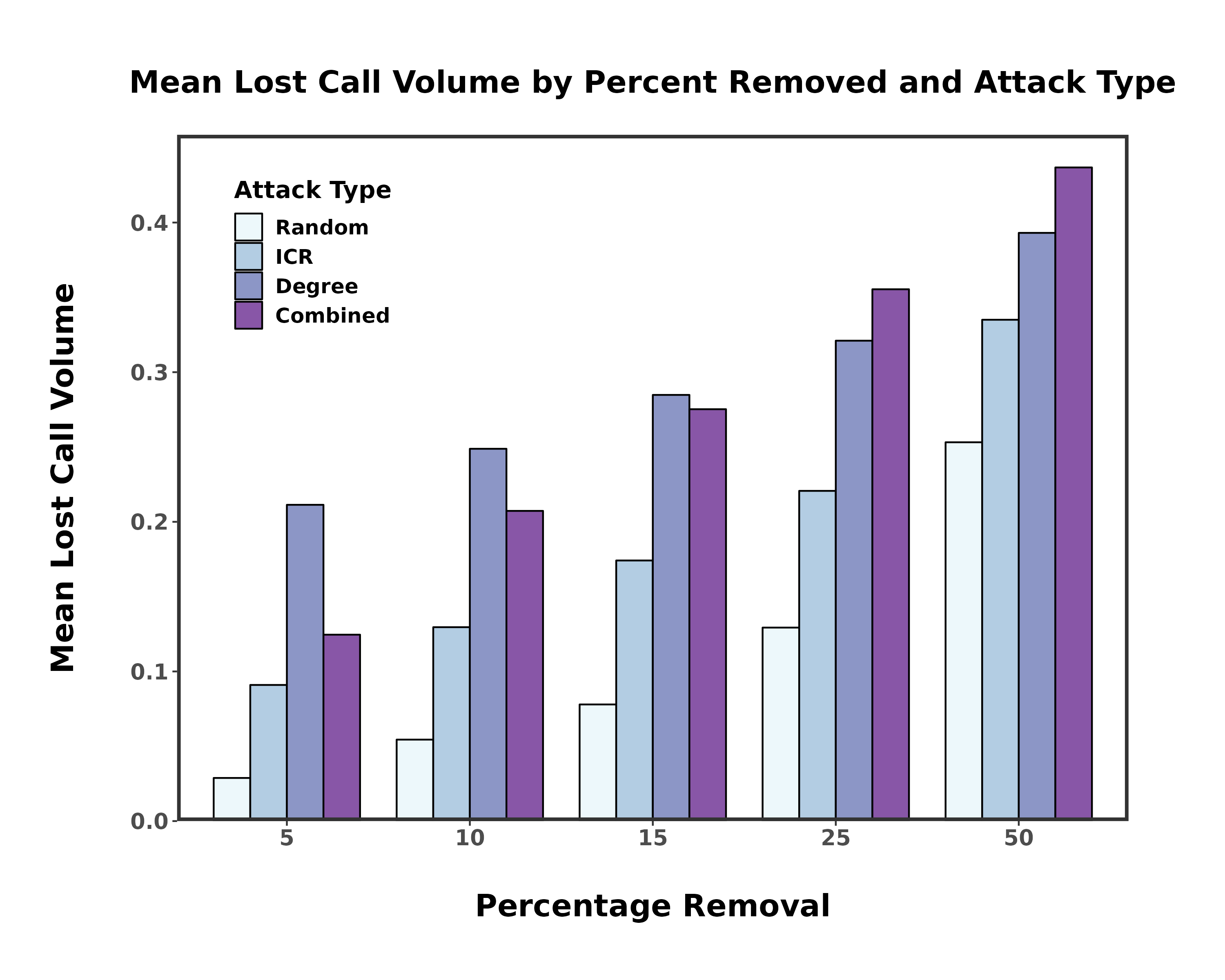}
\end{center}
\caption{Mean fraction of all calls directed to incapacitated nodes, by attack mechanism and removal fraction.  Targeted attacks lead to substantially greater loss of call volume than random attacks at all levels of removal, with removal of high-degree nodes generating greater losses than removal of ICRs per se. \label{fig:meanloss}}
\end{figure}

While the networks may appear to coalesce, this does not mean that they are more or less efficient than the baseline configurations. We know that these networks experience some level of call loss, but the attack type and percentage killed may lead to more or less call loss overall. We thus begin by examining the mean lost call volume, i.e. the proportion of all communication within a simulation devoted to ``calling the dead.'' Figure~\ref{fig:meanloss} shows the mean value over all of the communication networks for each removal strategy by percentage and attack type. First, it is evident that each of these attacks result in more calls to incapacitated nodes, on average, as we increase the number of nodes removed. We can also see that random attacks on average do not induce the same levels of call loss compared to the other attack types. Degree attacks at lower percentage of removal tend to be the most high-impact removal strategy, but this becomes slightly outpaced by the combined attack at much higher removal levels (25\% and 50\% removal). 

This is indicative of findings we might expect to see on \emph{a priori} grounds, especially with the behavior of random removal.  As observed in prior work \citep{petrescu-prahova.butts:ijmed:2008,renshaw:networksci:2023}, the WTC networks are strongly hub-dominated. These hub-structures are composed of  high degree individuals, and are frequently individuals who have been identified in pre-disaster contexts as being in an Institutionalized Coordinative Roles (ICR). When such individuals are removed, the same social mechanisms that fostered the initial emergence of hub roles (conversational inertia, preferential attachment, and ICR-directed communications) encourage persistence of attempts to contact them; by contrast, randomly selected responders are unlikely to occupy hub roles, and hence less likely to be persistent targets.  While non-response from incapacitated nodes will eventually reduce or extinguish this behavior, it takes time for this to occur (as we show below).

\subsubsection*{Specialists Waste More Calls on The Dead}

\begin{figure}
\begin{center}
\includegraphics[width=\textwidth]{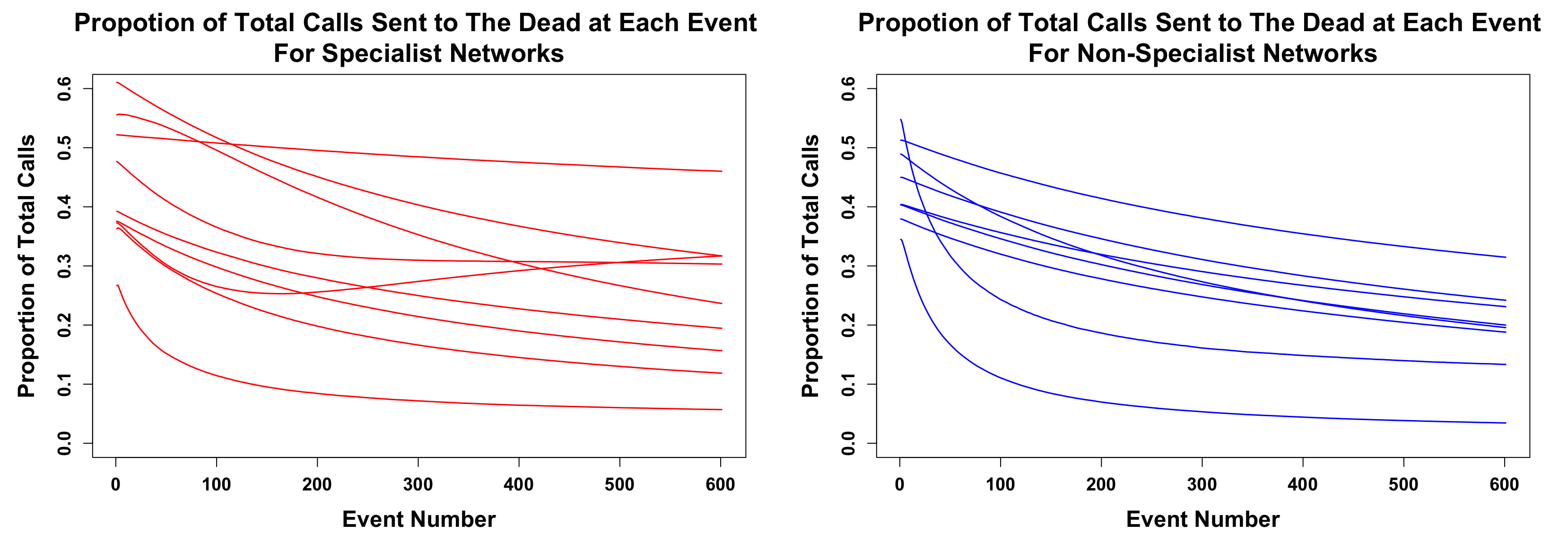}
\caption{Average call loss curves for specialist (left) and non-specialist (right) networks across all attacks.  While rates of calls to incapacitated nodes fall in nearly all networks, learning is faster on average in non-specialist networks (with some specialist networks showing high persistence in attempting to contact incapacitated nodes). \label{fig:crv.loss}}
\end{center}
\end{figure}

As was found in prior work relating to the WTC radio networks -- particularly in the recent work by \citet{fitzhugh.butts:sn:2021}, there are differences in efficiency between specialist and non-specialist groups. First, we test the mean differences in call loss percentage between the specialist and non-specialist networks, which can be seen in Table~\ref{tab:ttest_spec_loss_atk}. Results here show that there are statistically significant differences between the mean call volume loss between specialist and non-specialists, indicating that specialists across all attack mechanisms are more likely to contact inactive nodes.  

\begin{table}[h]
\setlength{\tabcolsep}{8pt}
\centering
\begin{tabular}{lcccccc}
  \hline
 & $t$-value & $p$-value & Mean Spec & Mean Non-Spec & Difference \\ 
  \hline
Combined & 10.13 & $p<$0.001 & 0.30 & 0.26 & 0.04  \\ 
  Degree & 11.88 & $p<$0.001  & 0.31 & 0.27 & 0.04  \\ 
  ICR & 24.46 & $p<$0.001  & 0.23 & 0.14 & 0.09  \\ 
  Random & 4.52 & $p<$0.001  & 0.12 & 0.10 & 0.01  \\ 
   \hline
\end{tabular}
\caption{Mean percent call loss (proportion of calls sent to removed/incapacitated alters) compared between Specialist and Non-Specialist networks.\label{tab:ttest_spec_loss_atk}}
\end{table}

On average, specialists expend anywhere from 1\% to 9\% more call volume on trying to reach incapacitated nodes than their non-specialist counterpart networks. We also can see that ICR attacks are particularly disruptive to the normal functioning of the specialist networks, with specialists spending nearly 10\% more call volume on incapacitated nodes than non-specialists. 

We also tested for differences by fraction of nodes removed. In Table \ref{tab:ttest_spec_loss_perc}, we can see that all of percentages are significantly different between the specialist and non-specialist networks, affirming that specialist waste more calls across all percentages of node removal. Looking at table \ref{tab:ttest_spec_loss_perc}, we find that in the lower attack percentages there is less of a gap between groups, with 3 percentage points in both the 5\% and 10\% removal cases. As the amount of nodes removed increases, the gap between specialists and non-specialists widens with a difference of 5, 7, and 6 percentage points for the 15\%, 25\%, and 50\% removal cases, respectively.

\begin{table}[h]
\setlength{\tabcolsep}{8pt}
\centering
\begin{tabular}{rrrrrr}
  \hline
 & $t$-value & $p$-value & Mean Spec & Mean Non-Spec & Difference \\ 
  \hline
5\% & 10.97 & $p<$0.001 & 0.13 & 0.10 & 0.03  \\ 
  10\% & 9.19 & $p<$0.001 & 0.17 & 0.14 & 0.03  \\ 
  15\% & 12.36 & $p<$0.001 & 0.23 & 0.18 & 0.05  \\ 
  25\% & 14.24 & $p<$0.001 & 0.29 & 0.22 & 0.07  \\ 
  50\% & 12.46 & $p<$0.001 & 0.38 & 0.32 & 0.06  \\ 
   \hline
\end{tabular}
\caption{Mean percent call loss (proportion of calls sent to removed/incapacitated alters) compared between Specialist and Non-Specialist networks within each percentage removed. \label{tab:ttest_spec_loss_perc}}
\end{table}

\subsection{Functionality}
\subsubsection*{Forward Reachability Increases After Node Removal}

\begin{figure}
\begin{center}
\includegraphics[width=4in]{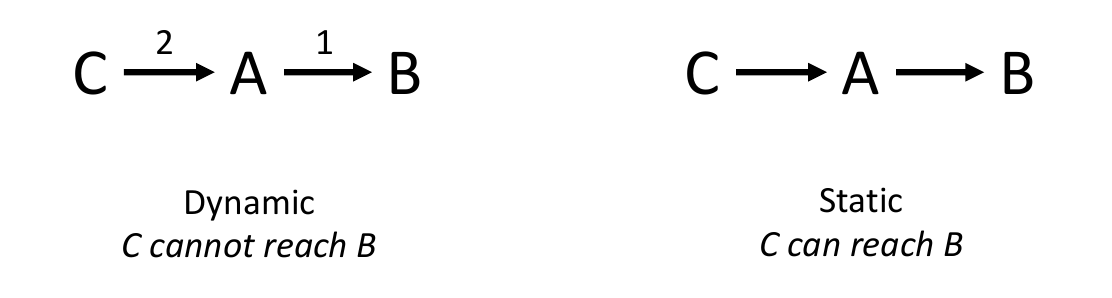}
\caption{Static versus forward reachability in a relational event system.  Node A sends to node B and then node C sends to node A. In a time-aggregated network, C can reach B because there is a directed path from C to B, via A. However, in the dynamic structure, C cannot reach B because it sent to A after A had already sent to B. \label{fig:reachEx}}
\end{center}
\end{figure}

While the ability of the of the WTC networks to reduce lost communication effort after attack can be determined by calls wasted on the dead, a general measure of functionality should capture the ability of nodes to pass information to others. Here, we assess this via a measure of forward reachability. We calculated forward reachability for each living node after the time of attack using the \texttt{networkDynamic} \citep{networkDynamic:2023} and \texttt{tsna} \citep{tsna:2021} packages for \textsf{R}. In particular, we measured what fraction of the network could be reached by a randomly chosen node, accounting for the time ordering of edges. While the time-aggregated network may be more connected after node removal (as was seen in our above analyses), this does not necessarily mean that nodes can actually reach each other in the dynamic network,  as relational events are directed and ordered. This difference between a static reachability measure and forward reachability is illustrated in Figure~\ref{fig:reachEx} 

This measure more accurately captures the functionality of the dynamic communication networks, and we can use it to assess the extent to which information passing potential is impaired by attack. Forward reachabilty was calculated for each node as the fraction of other ``living'' nodes it could reach and then averaged across all nodes in the network.

\begin{figure}
\begin{center}
\includegraphics[width=\textwidth]{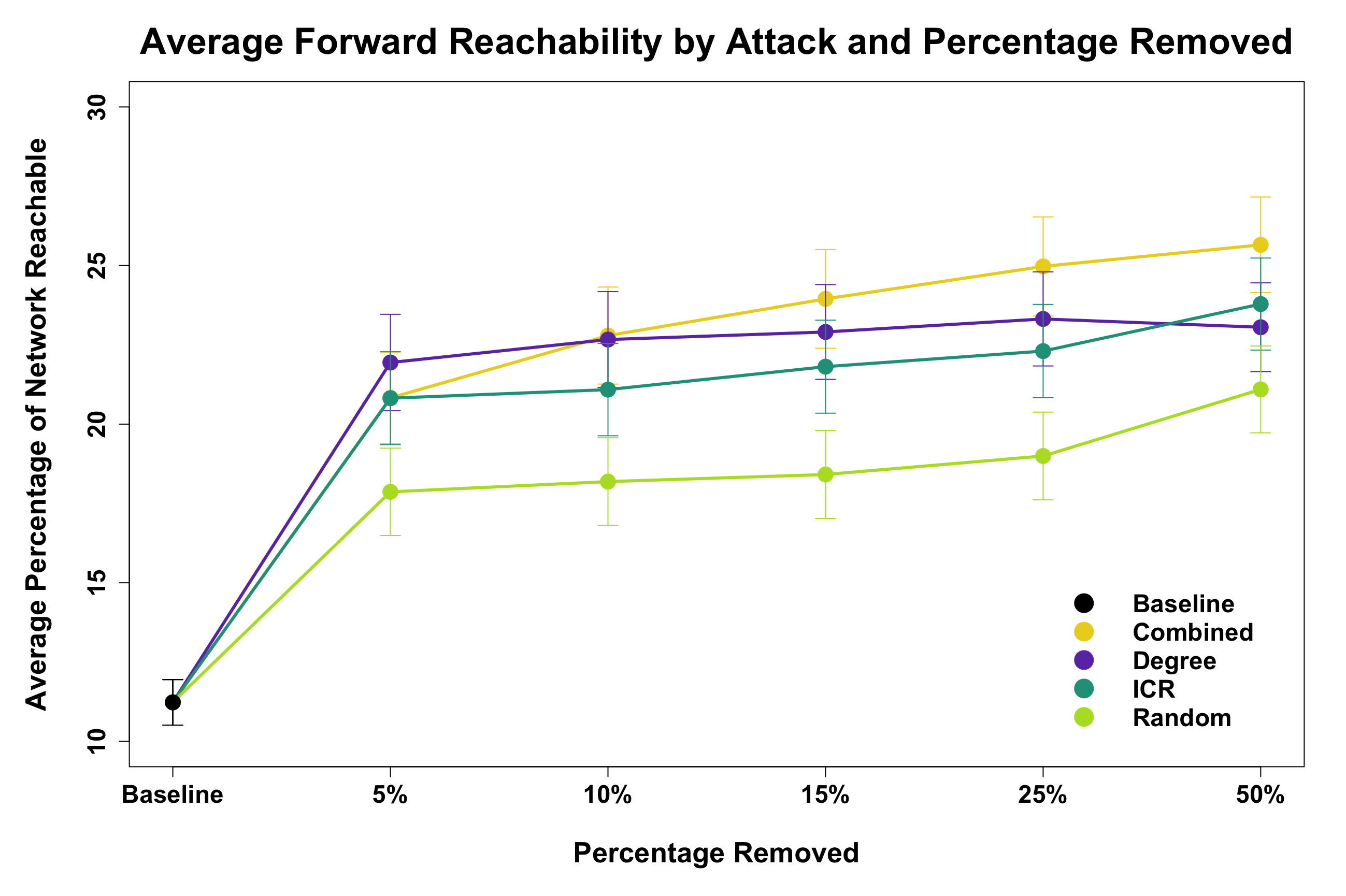}
\caption{Mean forward reachability by both attack type and percentage removed. Forward reachability tends to increase with the percentage of nodes removed, but at varying rates per each attack type. \label{fig:reachPlot}}
\end{center}
\end{figure}

Testing for differences between all simulations and baseline, we find that reachability is not only sustained post node removal, it almost doubles, with a given node being able to reach 11.2\% of the network on average in control versus 21.8\% when averaging over all treatments. When looking at how this varies by attack type, we find that combined and degree attacks are not statistically different and increase reachability the most compared to baseline. ICR attacks appear to cause a larger increase in reachability than random attacks, indicating that targeted attacks increase reachability more than non-targeted ones. More specifically, Combined and Degree attacks increase the forward reachability measure by 104.62\% and 107.22\% above baseline, respectively, ICR attacks increase forward reachability by 83.46\%, and finally random attacks cause an average increase of 77.91\%. 

In terms of the percent of actors removed, we find that all percentages removed are significantly different from one another. It appears that the increase to reachability scales positively with the number of actors removed, from 68.97, 77.76, 85.93, 99.42, and 134.44 for 5\%, 10\%, 15\%, 25\%, and 50\% respectively. As this effect is likely partially influenced by size, we also tested for differences in the average count of actors reachable by a given node. Despite the networks shrinking after node removal, reachability still demonstrates a statistically significant increase compared to baseline, with the differences between attack type and percentage removed remaining consistent. 

\begin{table}
\setlength{\tabcolsep}{8pt}
\centering
\begin{tabular}{rcccc}
  \hline
 Attack Type & Special. Average & Non-Spec. Average & lower CI & upper CI \\  
  \hline
Combined & 103.21 & 111.73 & -12.43 & -4.61 \\ 
  Degree & 97.53 & 112.62 & -19.04 & -11.13 \\ 
  ICR & 85.89 & 80.73 & 1.46 & 8.86 \\ 
  Random & 74.70 & 81.51 & -11.09 & -2.54 \\
  \hline
  Percentage Removed & Special. Average & Non-Spec. Average & lower CI & upper CI \\
  \hline
  5\% & 67.55 & 70.57 & -6.86 & 0.81 \\ 
  10\% & 77.46 & 78.10 & -4.73 & 3.45 \\ 
  15\% & 87.33 & 84.37 & -1.28 & 7.20 \\ 
  25\% & 98.66 & 100.28 & -6.05 & 2.80 \\ 
  50\% & 120.67 & 149.92 & -34.22 & -24.27 \\ 
   \hline
\end{tabular}
\caption{Change in mean forward reachability as a percentage of baseline compared between Specialist and Non-Specialist networks by percentage removed and attack type. \label{tab:ttest_spec_reach}}
\end{table}

Finally, when we compare the specialist and non-specialist networks, we find that specialist networks see a smaller increase in reachability compared to non-specialist networks. Combined attacks increased forward reachability, with specialists seeing a 103.21\% increase to their forward reachability (relative to control) and non-specialists seeing an increase of 111.73\%. For Degree-targeted attacks, the mean increase in reachability was 97.53\% for specialists, and 112.62\% for non-specialists. Random attacks increased baseline reachability by 74.7\% and 81.51\% respectively, and ICR attacks impacted specialists slightly more at 86\% compared to 80.7\% for non-specialists.

When we break out the information for difference in reachability by the percentage removed, we find few statistically significant differences, with the exception being 50\% removal; non-specialists increase reachability 150\% above baseline compared to 120.67\% for specialist networks. The differences between specialist and non-specialist networks were also confirmed in our test of the reachability count measure. 

\subsubsection*{Reserve Use Increases After Node Removal}

Prior literature on resilience in social insect colonies points to the use of reserves to maintain functionality and respond to crises. Knowing that our networks appeared similarly resilient in the case of node removal, we tested for the use of reserves as well. We defined reserves as those individuals who had not sent communications at the time of node removal, and who were not removed from the network. We then measured the proportion of these specific nodes that sent communications by the end of the simulation process, and defined this as the rate of \emph{reserve use}. We compare this measure to baseline to see how the removal strategy types and percent of actors removed affects the involvement of reserves. 

\begin{figure}
\begin{center}
\includegraphics[width=\textwidth]{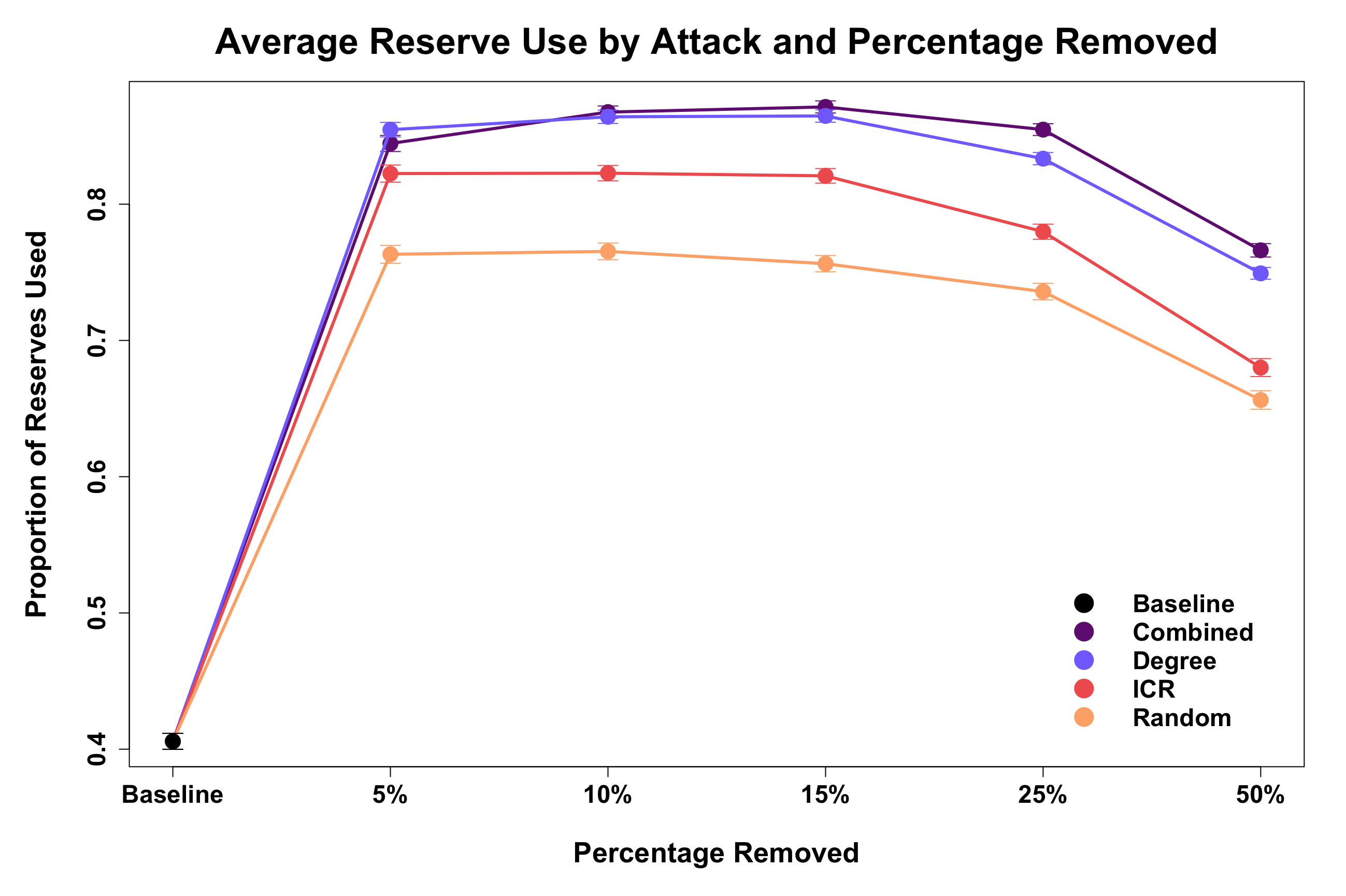}
\caption{Mean reserve use by both attack type and percentage removed. Reserve use is higher for all percentages removed, compared to baseline, but decreases at higher rates of node removal compared to 5\%-15\% removal. This pattern holds for all attack types, but at varying magnitudes. \label{fig:resUse}}
\end{center}
\end{figure}

Overall, networks under node removal conditions use an average of 80\% of their reserves, compared to only 40.57\% in baseline. Broken down by attack type, combined and degree attacks are not statistically different from each other, with reserve use increasing at 114\% and 113\% of baseline respectively. ICR attacks show an increase of 99.5\% above baseline compared to only 88.5\% for random attacks, indicating that targeted attacks are better for mobilizing reserves than random attacks, but that any node removal increases reserve use. 

When broken out by the percentage removed, we find that 10\% and 15\% removal are the only groups that are not statistically different and that reserve use increases from 109\% in the 5 percent removal case to 111.8\% for both 10 and 15 percent removal cases. Reserve use then shrinks to 104.6\% and 82.46\% in the 25 and 50 percent removal cases, respectively. 

\begin{table}[ht]
\setlength{\tabcolsep}{8pt}
\centering
\begin{tabular}{rcccc}
  \hline
 Attack Type & Special. Average & Non-Spec. Average & lower CI & upper CI \\ 
  \hline
Combined & 102.57 & 127.38 & -26.71 & -22.89 \\ 
  Degree & 103.48 & 124.66 & -23.38 & -18.97 \\ 
  ICR & 92.18 & 107.78 & -17.58 & -13.61 \\ 
  Random & 76.56 & 102.00 & -27.61 & -23.27 \\
  \hline
  Percentage Removed & Special. Average & Non-Spec. Average & lower CI & upper CI \\
  \hline
  5\% & 98.06 & 121.38 & -25.53 & -21.10 \\ 
  10\% & 102.74 & 122.05 & -21.63 & -16.98 \\ 
  15\% & 102.05 & 122.63 & -22.98 & -18.18 \\ 
  25\% & 91.52 & 119.35 & -30.11 & -25.55 \\ 
  50\% & 74.11 & 91.85 & -20.06 & -15.42 \\ 
   \hline
\end{tabular}
\caption{Mean reserve use as a percentage of baseline compared between Specialist and Non-Specialist networks within each percentage removed and attack type. \label{tab:ttest_spec_resUse}}
\end{table}

\begin{figure}
\begin{center}
\includegraphics[width=\textwidth]{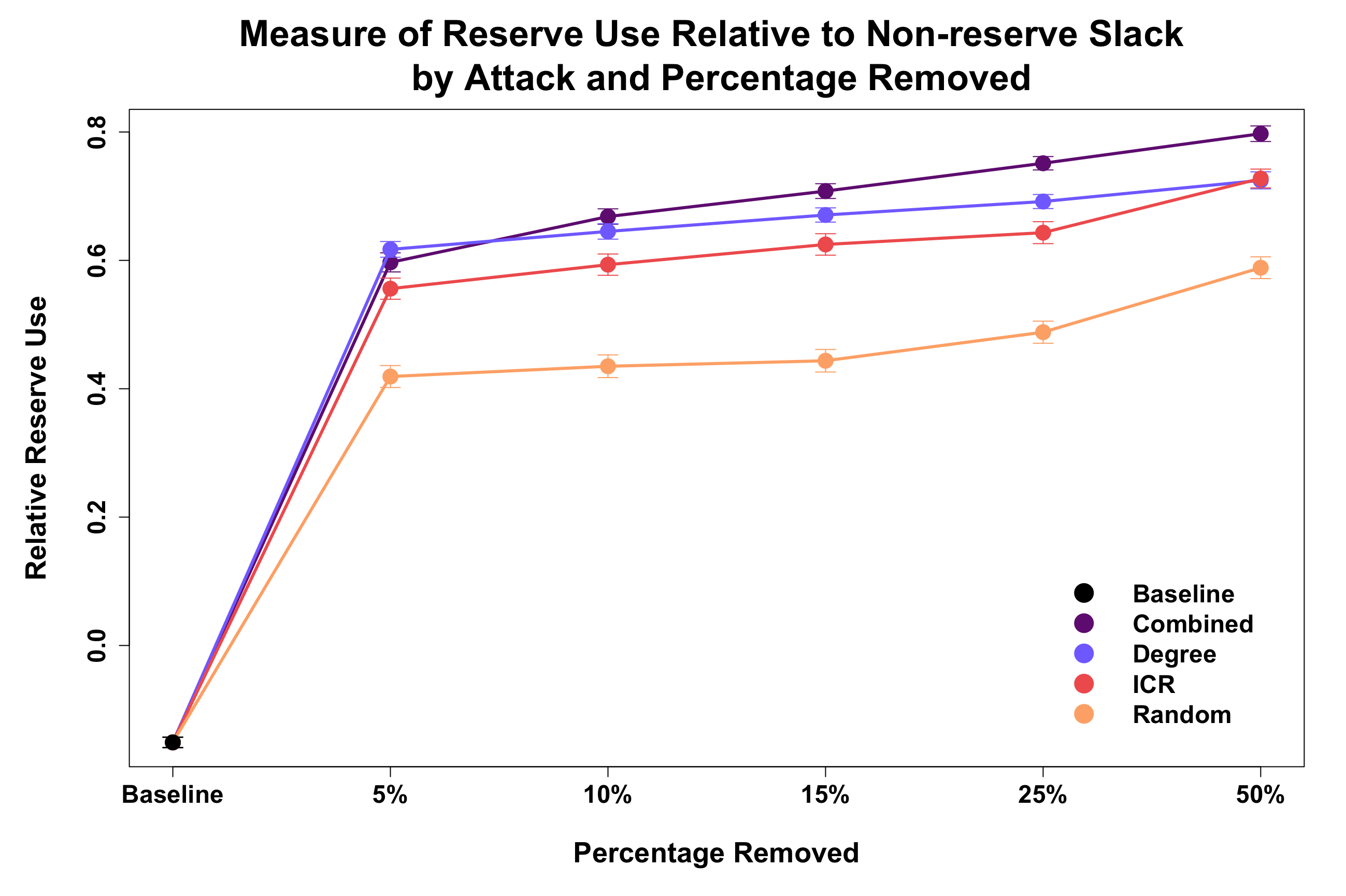}
\caption{The mean difference between the fraction of inactive nodes that become active, and the fraction of formerly active nodes that become inactive.  If $n_a$ is the count of previously active nodes, of which $s_a$ are later inactive, and if $n_i$ is the count of previously inactive nodes, of which $s_i$ are later active, this measure is then $\tfrac{s_i}{n_i} - \tfrac{s_a}{n_a}$. All attack types and percentages of nodes removed result in a positive measure, indicating that more reserves are being activated compared to the number of non-reserve nodes that are becoming inactive. We find the opposite pattern in our baseline simulations. \label{fig:resUseAlt}}
\end{center}
\end{figure}

Finally, for our specialized versus non-specialized tests, we find that non-specialized networks utilize reserves about 15 to 30 percentage points more under all node removal conditions. While reserve use clearly increases, there is a possibility that this is simply due to swapping places with active nodes, which is not quite consistent with how reserve use is conceptualized in the literature. In order to understand whether or not this increase was attributable to a decrease in non-reserve activity, we measured the fraction of inactive nodes that become active and compared it to the fraction of formerly active nodes that become inactive. This measure ranges from -1 to 1, with -1 indicating that no reserves were activated, while all formerly active nodes became inactive and 1 indicating that all reserves became active and no formerly active nodes became inactive. The results are shown in Figure~\ref{fig:resUseAlt} and indicate a positive measure for all attack types and node removal percentages. The baseline simulations are the only ones with a negative value indicating that more formerly active nodes are becoming inactive compared to the number of reserves that are becoming active. This measure is also positively associated with the percentage of node removal and all targeted attacks result in a lager measure than random attacks across percentages. Overall, this indicates that the increase in reserve activity is not attributable to formerly active nodes becoming inactive.   

\section{Discussion} \label{sec:discussion}

Our study provides novel insights into the resilience of human networks in the context of 17 radio communication networks during the unfolding World Trade Center disaster. We examined how these networks -- both specialist and non-specialist -- were able to cope with the incapacitation or death of key communicators, thereby providing valuable insights into their ability for dynamic response and resilience.

Arguably, the most striking finding of our simulation study is the observed tendency of the WTC networks towards increased coalescence after suffering personnel loss (particularly the loss of coordinators). While prior research within the network robustness paradigm has emphasized the potential for node removal to lead to fragmentation, our study suggests that human social networks may be more resilient to disruption than previously understood. Across the board, our networks demonstrate a higher rate of connectedness and involvement from actors in networks attacked by node removal. This tendency also coincides with not only a maintenance of functionality, but an \emph{increase} in it as measured by forward reachability.  Such effects could not have been seen in robustness studies (including prior robustness studies on the WTC networks), since by definition such studies do not allow for adaptation of the social system to disruption.  Although we note that our findings do not undermine those studies -- they are measuring different things -- they do suggest that care may be needed when \emph{interpreting} robustness studies in settings where adaptation to damage is likely.

Our approach also allowed us to identify notable differences in how two types of organizations would be expected to respond to death or incapacitation of members during a disaster event. Specialist networks displayed a greater difficulty in adapting to the loss of key actors by spending more time contacting the dead than their non-specialist counterparts, while also coalescing less and demonstrating a lower increase in forward reachability. This relative lack of adaptation appears to be a result of their dependence on Institutionalized Coordinator Roles (ICRs), potentially due to more rigid reliance on organizational standard operating procedures -- a factor identified in prior research as a potential constraint on flexibility in hazard and disaster contexts \citep{carley:os:1992}. In contrast, non-specialist networks demonstrated superior resilience and flexibility in response to the attacks, with their adaptability involving a greater mobilization of reserves as well as a greater increase in their network functionality with respect to forward reachability. 

Drawing parallels with social insects, organizational slack management, and engineering studies of node removal, this research highlights the critical role that reserve mobilization plays in the context of network resilience -- observing a general tendency across the board that previously non-participating individuals had increased activation post-removal. Not only did we note a decrease in the percentage of isolates in the networks when compared to baseline, but we find that the sending activity of those who are isolates directly following node removal increased more than in baseline networks.  While pressures for efficiency often suggest elimination of slack resources, our study reinforces the potential value of such resources in times of disruption. 

\subsection{Future Research}

Compared to most node removal studies conducted to date, we find that our networks do not have a tendency to fragment and disconnect upon death or incapacitation. Instead, we find that there is a general tendency towards what we call ``coalescence.'' Rather than fragmenting, coalescence is the ability for the network to restructure itself in a way that increases overall participation and connectivity while more evenly distributing the communication across actors; in our study this occurs in response to the removal of key communicators in targeted attacks. While this suggests the potential for higher levels of resilience in human social networks than has been implied by robustness studies, we observe that one should not conclude that such resilience is always optimal; for instance, the same adaptations that make specialist networks in the WTC less resilient in our study may make them more efficient when not experiencing personnel loss.  Likewise, desirability of network resilience is obviously in the eye of the beholder: resilience in criminal networks, networks of proliferating cells in a tumor, or the like may pose obstacles to social or health intervention.  Regardless, our findings suggest that researchers may need to give more consideration to network adaptation in interpreting node and edge removal studies. Probing the behavior of dynamic models to edge or node removal can account for potential temporal variability in how networks respond, resulting in considerably different outcomes than would be suggested by removal of elements from a static network. 

An obvious benefit of robustness studies is that they are simpler to perform than resilience studies, as one needs only a single empirical observation; a perhaps less obvious benefit is the lack of a need for assumptions about how a network will respond to damage.  This is an inherent tradeoff, since resilience is \emph{per se} a dynamic phenomenon.  To that end, it should be observed that an increasingly wide range of approaches (including but not limited to relational event models, temporal exponential family random graph models, and stochastic actor-oriented models) are available for developing and validating empirically calibrated models of network dynamics.  Such models offer considerable opportunity to broaden our understanding of resilience, not only in communication networks but in other types of systems.

It should also be observed that there are many other kinds of resilience studies that could be carried out. Here, we held the behavioral mechanisms governing the networks fixed, and saw how the networks responded to disruption. In some settings, attacks may trigger changes to network dynamics themselves. While this is something that we cannot examine here (since we have no data to use for calibration), this would seem to be a fruitful area for future empirical and theoretical research. One could also consider other types of interventions beyond those conducted here, including alternative removal schedules, testing the impacts of intermittent failures, or focusing on edge removal, or even edge ``scrambling'' so that the intended recipient is unaware of the targeted nature of the communication, as prior work has found that conversational norms like turn-taking are critical for phenomena such as hub formation \citep{gibson.et.al:po:2019}.

Finally, we highlight some practical implications of our findings for enhancing resilience in emergency management contexts. While we cannot speak directly to effectiveness of the resulting networks in terms of how they reorganize after being attacked, from a purely structural standpoint -- where individuals are capable of maintaining connectivity and reducing loss post-removal -- we find that there are some obvious gains to not being too dependent on specialized roles and actors in communication networks (such as ICRs in our case). Further, training that emphasizes the need to adapt to apparent losses (when an agent ``goes silent'') and switch to communications with other available agents, may expedite communication adaptation when loss occurs. This, of course, must be balanced against the gains in efficiency and communicative memory that is obtained by focusing coordination costs on a small number of (ideally trained and institutionally identified) individuals. We cannot speak to where this balance is, but we do show that there are costs to resilience from this strategy.

\section{Conclusion} \label{sec:conclusion}

To conclude, our study provides insights into the dynamics and resilience of both specialized and non-specialized networks in response to personnel loss, allowing us to see into the potential behavioral responses to a real-world context in an unfolding emergency. The observed shift toward coalescence following network disruptions as well as the potential rigidity of reliance on institutionalized coordinators highlights an intricate interplay between structural and behavioral factors in the dynamic resilience of the WTC networks.

Crucially, our findings suggest a need for a more nuanced interpretation of traditional approaches to studies of network robustness -- particularly as it applies to understanding network fragmentation.  While robustness studies are unquestionably valuable from a purely structural standpoint (i.e., in showing e.g. how connectivity is maintained within particular networks), their utility in predicting responses to damage in real systems depends on the often tacit assumption that damage will not be restored on the relevant timescale.  While this is often viewed as a convenient but reasonable approximation, this may always be accurate: in the case of the WTC networks, we see rapid and dramatic adaptation to personnel loss that results in \emph{enhancement} of connectivity, something that could not be seen from a robustness study.  Particularly where robustness studies are employed to inform policy or network design, it is important to rule out or otherwise account for such effects.  Relatedly, our findings suggest that factors that do not contribute to robustness (most prominently, the presence of minimally connected or even disconnected nodes in the original network) may contribute to resilience, as when previously uninvolved personnel become mobilized when demands escalate.  The utilization of reserves in the WTC networks suggests that the presence of ``social loafers'' or ``slack'' should to be considered in future studies as a potentially adaptive feature for resilience to personnel loss, as has been found in the social insect literature. 

Finally, our study highlights the value of recent advances in statistical network analysis for theoretical studies of social process, and for the ability to approach questions relating to the resilience of social networks in an empirically informed manner while still working within the often severe limits of available data.  Not mere hypothesis-testing tools, generative models for social process constitute formal theories for how the world unfolds, and can be employed for a wide range of purposes.

\bibliography{CTD}

\end{document}